\documentclass[conference,10pt,a4paper]{IEEEtran}

\ifCLASSINFOpdf
  \usepackage[pdftex]{graphicx}
\else
  \usepackage[dvips]{graphicx}
\fi
\usepackage[cmex10]{amsmath}
\usepackage{array}
\usepackage{eqparbox}
\usepackage{fixltx2e}
\usepackage{url}
\usepackage{siunitx}

\begin{document}

\title{Nanomechanics of CNTs for Sensor Application}

\author{

\IEEEauthorblockN{Christian Wagner\IEEEauthorrefmark{1}, Steffen Hartmann,\\Bernhard Wunderle\\
Center for Microtechnology\\Chemnitz University of Technology\\09107 Chemnitz, Germany}
\IEEEauthorblockA{
\IEEEauthorrefmark{1}Email: christian.wagner@zfm.tu-chemnitz.de}
\and
\IEEEauthorblockN{J\"org Schuster, Stefan E. Schulz,\\ Thomas Gessner\\
Fraunhofer Institute for\\Electronic Nano Systems\\ 09126 Chemnitz, Germany}
}

\maketitle

\begin{abstract}
A nanoscopic simulation for an acceleration sensor is aimed based on the piezoresistive effect of carbon nanotubes (CNTs). Therefore, a compact model is built from density functional theory (DFT), compared with results of molecular dynamics (MD) that describes the mechanics of carbon nanotubes in a parameterized way. The results for the interesting kind of CNTs [(6,3) and (7,4)] within the two approaches agree in a satisfying way, when DFT-calculations are performed with atomic configurations obtained by MD geometry optimization. Geometry optimization yields the Poisson's ratio for CNTs. Thus, values from MD and DFT are compared. The simulation finally aims the modeling of the conductive behavior of CNTs when strain is applied, but this needs further verification. Here, we present the prediction of the tight binding model for suitable CNTs.
\end{abstract}


% begin text here

\section{Introduction}

The trend of miniaturization in microelectronics also affects the world of sensors. The advantages of smaller sensors is obvious as integration into actual devices is easier and more space-saving. Another advantage is that one needs less material and so, the amount of natural resources e.g. rare earth metals, are preserved. Today, the nano scale is explored to look out for new sensing principles and one of them is the piezoresistive effect of carbon nanotubes (CNTs), useful for a nano-scaled acceleration sensor.

CNTs have outstanding mechanical and electrical properties -- they are very stiff (their Young's modulus is about 1 TPa for ideal CNTs, \cite{Wu_2008}), flexible (they can be elastically strained up to 10 \% \cite{Park_2010,HaiYang_2009}) and they show strong electrical response on mechanical load \cite{Maiti_2002,Minot_2003}. Their conductivity can change up to one order of magnitude while being strained only about 1-2\% due to bandgap change. The piezoresistive behavior strongly depends on the type of CNT used. 

CNTs can be considered as rolled up graphene sheets - the way, how they are rolled up, is given by their chiral indices (n,m). They can be classified into metallic, semiconducting and semimetallic ones, where the metallic (n,n) CNTs do not show any piezoresistivity, but the (n,0) (semiconducting or semimetallic) are most sensitive to mechanical load. The other classes show an intermediate sensitivity depending on their chiral angle\cite{Yang_2000,Kleiner_2001}. Chiral CNTs with integer $(n-m)/3$ are semimetallic and they best fit the needs for a sensor, because their bandgap in the relaxed state can be overcome by the energy of a charge carrier at room-temperature. This leads to a high conductance.

Experimentally fabricated CNTs contain a mixture of all the different types and type-selective growth techniques are still a big challenge. They can be chemically separated, but the result is a subset of all the types available. For realizing a sensor, it is planned to integrate CNTs with a tube-radius of about 1 nm. Looking out for semimetallic CNTs fitting this criterion and considering the data from the experimental groups, one can find the (7,4), (8,5) and (9,6) CNT as appropriate candidates. 

The modeling of the piezoresistive effect of CNTs using tight-binding models is already explored in literature. The most important issue is the change of bandgap on mechanical load, firstly predicted in \cite{Yang_2000,Kleiner_2001}. The approximations made in this model are not yet fully confirmed by a more fundamental theory, namely density functional theory (DFT). Tight binding theory only assumes nearest-neighbor coupling of localized electronic states. Thus, coupling of opposite carbon atoms in the tube cannot be included -- however these radius effects are the reason for the tiny bandgap of semimetallic CNTs. Besides the effect on the bandgap, the negation of many-body-effects in the tight-binding model can lead to changes in bandstructure. These effects are important for electronic transport calculations and thus for the size of the sensor signal.

Simulation of the above mentioned kinds of CNTs are rarely found in literature due to their big unit cell (124 atoms for the (7,4), 172 for the (8,5) and 228 for the (9,6) CNT). The (7,4) CNT has the biggest unit cell that DFT can deal with in a reasonable computation time. DFT has a typical $O(N^3)$ scaling behavior where $N$ is the number of valence electrons within the setup \cite{Mattsson_2005}. The (6,3) CNT with 84 atoms in the unit cell shows about the same properties as the other, semimetallic ones do and its mechanical properties have also been calculated in literature \cite{Bogar_2005}. Thus, it is a good reference. 

Since the real structure of the CNTs determines its electronic structure, the first step towards a reliable sensor model is to understand the mechanics of carbon nanotubes in detail before understanding the electronics. This task could be fulfilled by molecular dynamics simulations (MD), but there, electronic structure of the CNTs and thus the band structure cannot be derived. That is the reason why we also explore the mechanics by means of ab-initio methods. 

The simulation of the entire sensor element will be achieved by using the framework of the virtual hardware description language (VHDL-AMS) that uses lumped elements to describe sub-elements performing distinct tasks. The characteristic of each lumped element is usually explored by underlying, classical physics simulations, e.g. solving coupled partial differential equations for all the relevant physical phenomena. In difference to these classical simulations, we incorporate a compact nano model based on quantum mechanical ab-initio simulations to introduce it into the device simulation.

\section{Calculation details}

For the simulation of the carbon nanotubes, Atomistix ToolKit (ATK) from QuantumWise is used \cite{Quantumwise_2011,Brandbyge_2002}. Within the zoo of available exchange functionals, the local density approximation (LDA) is the simplest one with the lowest computational cost being sufficient for pure carbon systems. The Perdew-Wang version of this functional has been applied\cite{Perdew_1991}. For geometry optimization of the nanotubes, we cut off, when the remaining forces fall below 0.01\,eV/\AA{}. 

In the case of molecular-dynamics simulations, a Tersoff-Brenner-Potential for Carbon \cite{Brenner_2002} is used. The optimization terminated, when the relative change of the total energy fell below $10^{-16}$ and the remaining forces were lower than \num{10e-4} $10^{-4}$\,eV/\AA{}.

\begin{figure}[!t]
\centering
\includegraphics[width=\linewidth]{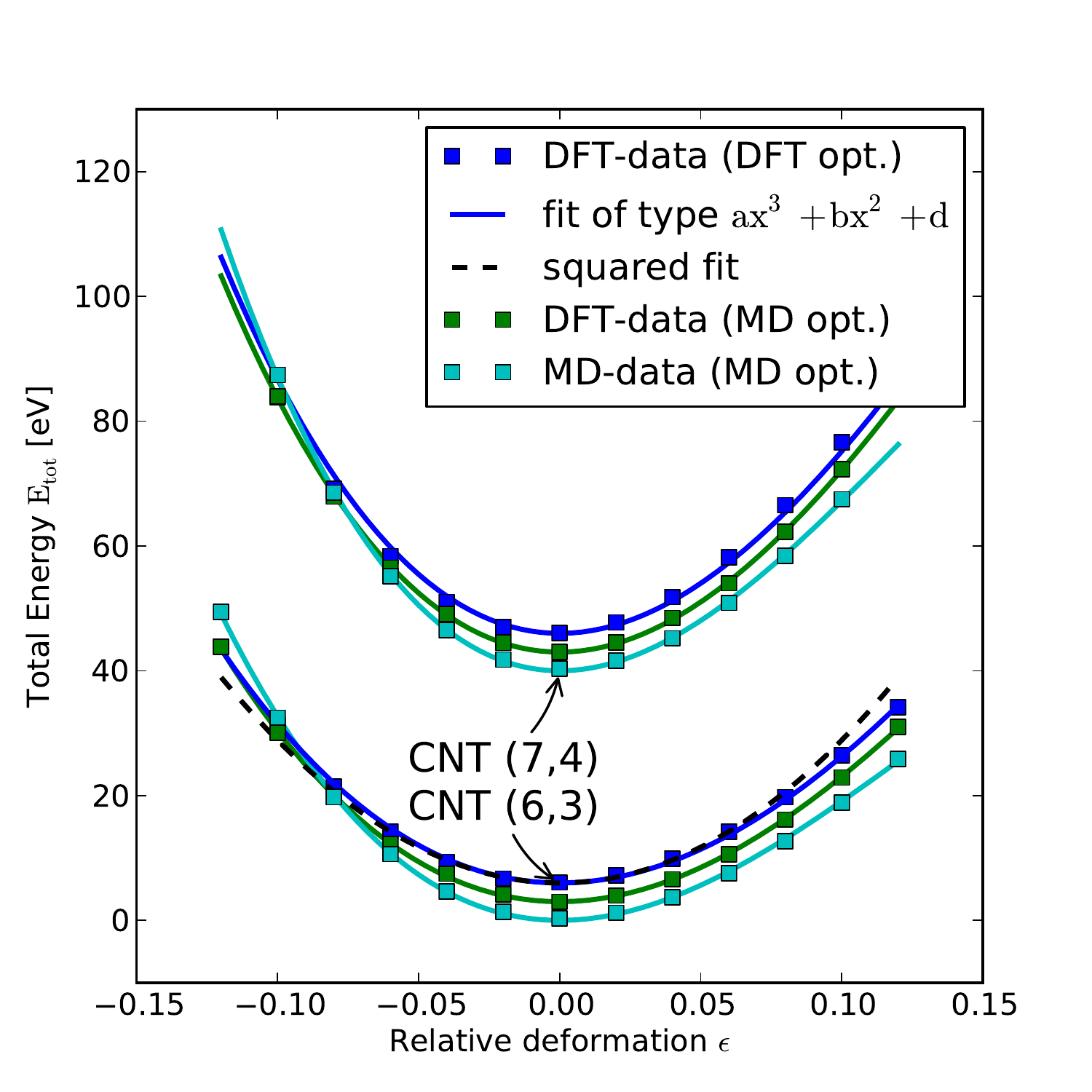}
\caption{The total energy of CNTs over strain: blue points represent DFT results, the green ones DFT results with atomic coordinates from MD and the cyan one MD results. The black dottet line is a squared fit that does not describe the data over the whole range. This is the reason for the third-order polynomial fit (solid line). Data are shifted vertically for a better visualization (details: see text).}
\label{fig_TotalEnergy}
\end{figure}

For the calculations of the nanotubes, we used periodic boundary conditions (PBC). The mechanical behavior is obtained by tracing the total energy stored in the (un-)strained atomic configuration and extracting mechanical parameters like force and Young's modulus. 

\section{Results}

There exist several publications on the Young's modulus of CNTs with a nice collection in \cite{Wu_2008}. Our calculation give values in the range of about 1 TPa as shown in table \ref{table_YoungModuli}. They slightly depend on the kind of CNT and whether data points from the compressive or tensile region are included to fit the total energy of the system. Those in table \ref{table_YoungModuli} are retrieved from a cubic fit up to $\pm$ 10\% to determine the Young's modulus to give comparable data to the current literature. It has been extracted by the coeffcient of the squared term. The parameters shown in table \ref{table_PolyParam} are fit taking into account the tensile region only, because there, the simulation results are meaningful for the sensor. Consequently, both tables cannot be compared directly.

In figure \ref{fig_TotalEnergy}, the total energy of a (6,3) and a (7,4) CNT is shown. The data points represent the DFT or MD results while the lines are polynomial fits of the kind $E_\text{tot}(\varepsilon) = a\varepsilon^3+b\varepsilon^2+d$. $\varepsilon$ denotes the relative strain. The different curves have different offset energies as the total energy of a system itself has no practical meaning. Only the energy difference is physically important. Thus, the energies of the different CNTs have been shifted by 40\,eV while the different energies of the same CNT are shifted by 3\,eV for a better visualization. The colors indicate different methods: DFT results (blue), MD results (cyan) and DFT calculation of geometries obtained by MD (green). 

\begin{table}[!t]
\renewcommand{\arraystretch}{1.3}
\caption{Young's moduli of different CNTs}
\label{table_YoungModuli}
\centering
\begin{tabular}{c|cc}
\hline
\textbf{CNT} & \textbf{DFT [GPa]} & \textbf{DFT after MD [GPa]} \\
\hline
(6,3) & 976 $\pm$ 14 & 997 $\pm$ 7\\
(7,4) & 991 $\pm$ 14 & 1012 $\pm$ 7\\
\hline
\end{tabular}
\end{table}

\begin{table}[!t]
\renewcommand{\arraystretch}{1.3}
\caption{Coefficients for $E_\textnormal{tot}(\varepsilon)=a\varepsilon^3+b\varepsilon^2+d$}
\label{table_PolyParam}
\centering
\begin{tabular}{l|cc}
\hline
\textbf{CNT} & \textbf{$a$ [eV]} & \textbf{$b$ [eV]} \\
\hline
(6,3) DFT & -4390 $\pm$ 150 & 2472 $\pm$ 22\\
(6,3) DFT/MD & -3090 $\pm$ 230 & 2300 $\pm$ 40\\
(6,3) MD & -3690 $\pm$ 30 & 2214 $\pm$ 6\\
(7,4) DFT & -6550 $\pm$ 240 & 3500 $\pm$ 40\\
(7,4) DFT/MD& -3650 $\pm$ 600 & 3290 $\pm$ 80\\
(7,4) MD& -5127 $\pm$ 28 & 3216 $\pm$ 6\\
\hline
\end{tabular}
\end{table}

\begin{figure}[!t]
\centering
\includegraphics[width=\linewidth]{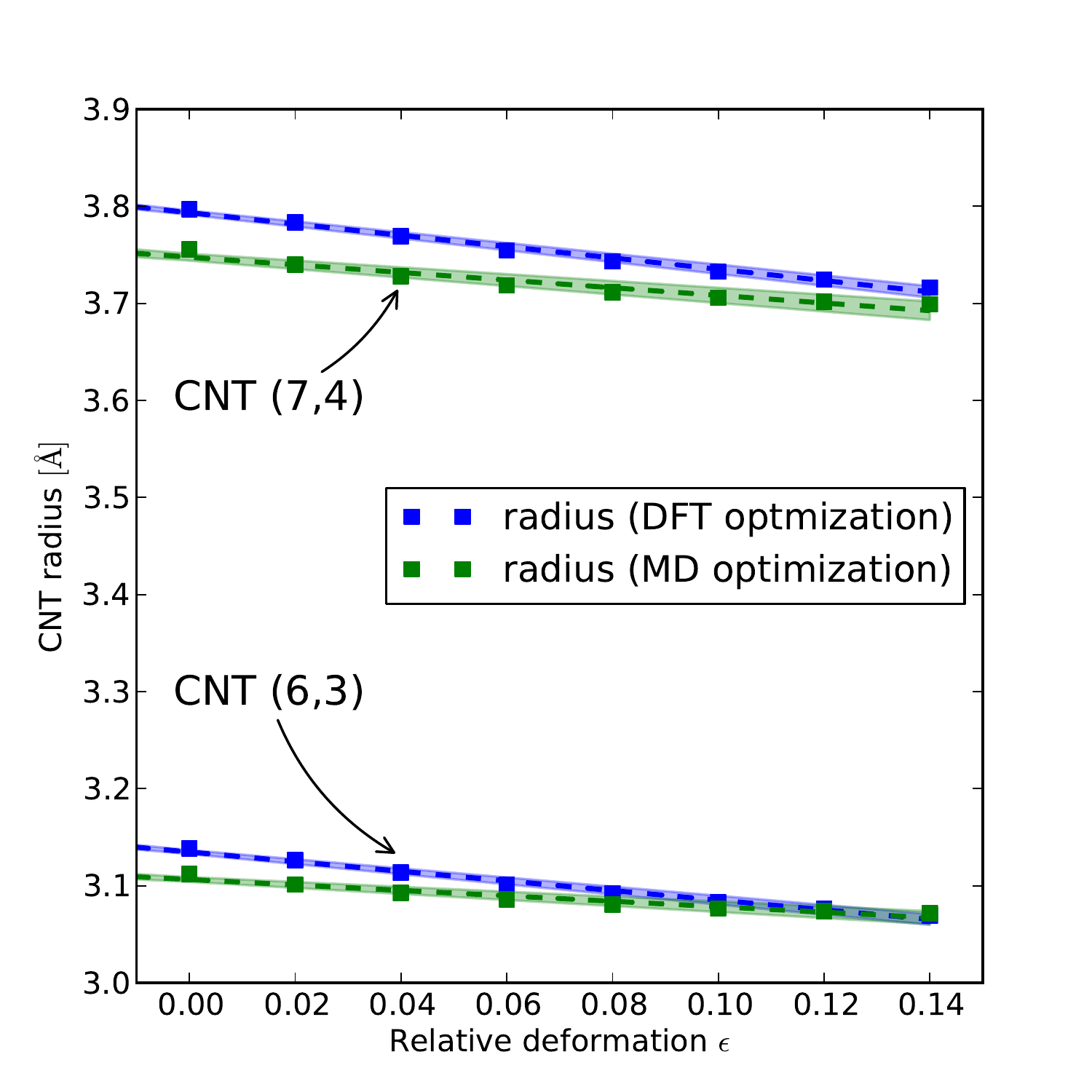}
\caption{The radius of (6,3) CNT and (7,4) CNT over strain: The blue lines show results from DFT optimization and the green ones are obtained from MD. MD generates in general tubes with a smaller radius and DFT optimization. Errorbars are not visible, because almost all carbon atoms lie on almost the same radius ($\Delta r\approx 0.001$\,\AA{}).}
\label{fig_RadiusStrain}
\end{figure}

Besides the comparison of the methods, which is convenient in itself, DFT calculations on MD geometries have been performed. The good agreement of those calculations (blue and green data points) in comparison to the deviations by pure MD results (cyan data points) shows that the DFT total energy does only slightly depend on the concrete atomic configuration. Mechanical properties of the CNTs are therefore insensitive to minor geometrical changes. The difference between the geometric configuration -- MD predicts a smaller radius than DFT -- is shown in fig. \ref{fig_RadiusStrain}, discussed later on. DFT leads to more trustworthy results for CNTs than MD, because it describes delocalized electron wave functions in a better way. MD only thinks in terms of atomic neighbor-interaction leading to about 1\% smaller C-C bond lengths than DFT. 

The final goal is to do the geometrical optimization with MD and to use DFT for electronic structure computations, only. In contrast to the mechanical properties, the, the electronic structure itself sensitively depends on the geometric configuration, so this method cannot be applied, yet. An important task is to bring these methods into agreement or to modify the MD geometries in a way that it is close to DFT optimal configuration. Then, a computationally cheap DFT-optimization can be run.

As one can also see in figure \ref{fig_TotalEnergy}, a squared polynomial fits (black, interrupted line) the data only up to $\pm$ 5\% and one finds that the total energy is asymmetric for higher strains. A similar result for another kind of CNT has been found by \cite{Valavala_2008}. That is why a third order polynomial is chosen to describe the the data correctly over the whole simulated range. The linear term was neglected, because at zero strain, the CNT is fully relaxed and in its energetic minimum. This third order term is even a more fundamental term, already described in literature about nanomechanics \cite{Colombo_2011,Shodja_2011}.

\begin{figure}[!t]
\centering
\includegraphics[width=\linewidth]{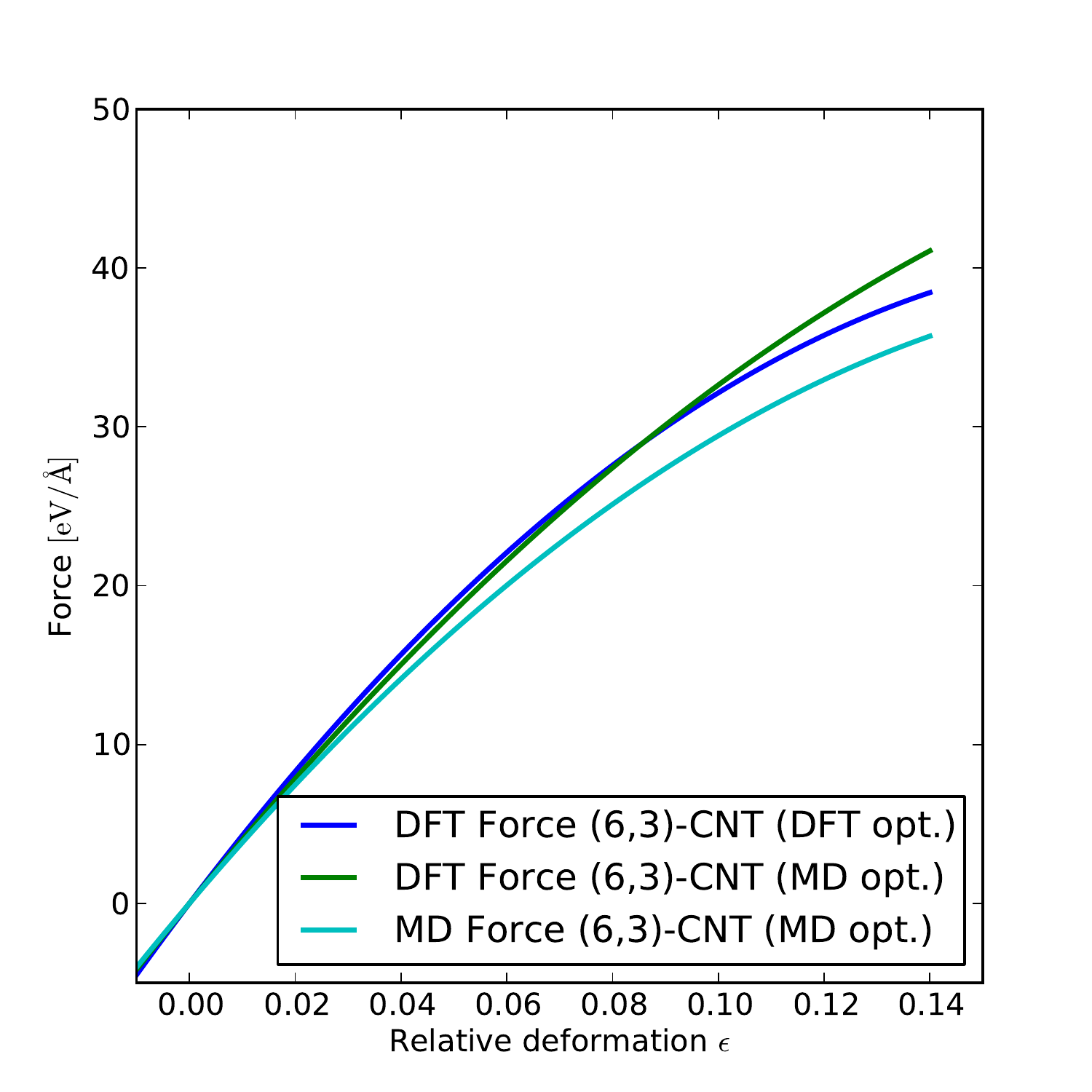}
\caption{The forces obtained by DFT (blue), MD (cyan) and DFT results of MD data (green) is shown. The curves are obtained from the fit of the tensile strain. The pure MD results in comparison to the others lead to slightly softer CNTs. The other results (DFT/DFT+MD) agree up to 10 \% strain.}
\label{fig_Force}
\end{figure}

The polynomial coefficients $a$ and $b$ are listed in table \ref{table_PolyParam}. $a$ is fluctuating strongly, because the third order coefficient is more sensitive on the detailed fitting procedure than the second order term. The latter used to extract the Young's modulus. The offset $d$ does not matter the mechanical properties, so it is not listed. The physical meaning can be obtained by calculating the forces shown in figure \ref{fig_Force}:

\begin{subequations}\label{eq_mech}
\begin{align} 
F(\varepsilon) [\text{eV/\AA{}}]&= -\frac{1}{\ell_0}\frac{\partial E_\text{tot}(\varepsilon)}{\varepsilon} = \frac{-3a}{\ell_0}\varepsilon^2 + \frac{2b}{\ell_0}\varepsilon \label{eq_Force}\\ 
E [\text{eV/\AA{}}^3]&=\frac{1}{A_0}\left.\frac{\partial F}{\partial \varepsilon}\right|_{\varepsilon=0} = \frac{2b}{\ell_0A_0} \label{eq_Young}
\end{align}
\end{subequations}

The Young's modulus $E$ is the linear dependency of the stress on the strain and thus is defined by equation  \ref{eq_Young}, where $\ell_0$ is the length of the CNT unitcell and $A_0$ is the cross section area of the CNT, defined by $A_0=2\pi rr_\text{VdW}$ with the tube radius $r$\,[\AA{}] and the according Van-der-Waals radius $r_\text{VdW} = 3.4$\,\AA{} \cite{Jeong_2010,Lu_1997}. 

The coefficients shown in table \ref{table_PolyParam} are obtained from fitting the positive direction (tensile strain), which is the application range within the compact model. Futher, the derived forces are shown in fig. \ref{fig_Force} -- the MD force is in general smaller than the force obtained by DFT due to the different approach. Both DFT calculations fit up to 10\% tensile strain and then, deviations occur. This is due to smaller deviations in some energy points that lead to stronger deviations in the fitted curve.

The compressing region is not taken into account, because the CNT used in the sensor has a high aspect ratio $\ell/d$ (CNT length $\ell \approx 1\,\mu$m, diameter: $d\approx 2\,$nm) of about 500 and will be bent upon compression, as a lot of MD-simulations have already shown \cite{Feliciano_2011,Jackman_2011,Nikiforov_2010b}. We partly observe shell-buckling that is also found in literature \cite{Feliciano_2011}, but this result is not reliable, since the DFT-modeled CNTs consist of one unit cell. Nevertheless, the compressed region is simulated within DFT. By this, a numerically stable fit of the Young's modulus is obtained allowing the comparison to other literature. 

Besides the Young's modulus, the Poisson's ratio plays an important role for the mechanics in general and it should be reproduced within our calculations. Often, the moduli are assumed to be in the range of 0.15 to 0.2 \cite{Bogar_2005}. In this reference values for all kind of thin CNTs with the chiral indices starting from (3,1) and ending up in (6,6) are reported. 

As it can be seen from figure \ref{fig_RadiusStrain}, the radius shows a linear dependence on the strain considering only the tensile region. Deviations of the atomic coordinates to the average radius are that small that errorbars in this figure are not visible. Thus, one can obtain the Poisson's ratio using both methods, which are shown in table \ref{table_Poisson}. DFT yields CNTs with a bigger radius than those from MD and also showing a higher radial contraction -- consistent with the results shown in \cite{Bogar_2005}. The corresponding magnitude in a basal-plane of graphite is 0.16 (\cite{SanchezPortal_1999} and references therein) making DFT results more trustworthy than MD results at this point.

\begin{table}[!t]
\renewcommand{\arraystretch}{1.3}
\caption{Poisson's ratio $\nu$ for (6,3) and (7,4) CNT}
\label{table_Poisson}
\centering
\begin{tabular}{l|cc}
\hline
\textbf{CNT} & \textbf{DFT} & \textbf{MD} \\
\hline
(6,3) DFT & 0.158$\pm$ 0.008 & 0.090$\pm$0.010\\
(7,4) DFT & 0.153$\pm$ 0.007 & 0.105$\pm$0.011\\
\hline
\end{tabular}
\end{table}

The next step is the calculation of the electronic structure that is needed to get information about the conductivity. The simplest model for the conductivity within the ballistic transport regime is to evaluate the bandgap and assume thermal activated carrier generation, which ends up in the following relationship of bandgap $E_G$ and resistivity $R$ with a given contact resistance $R_C$ \cite{Maiti_2002}:

\begin{equation} \label{equ_resistivity}
R = R_C + \frac{1}{|t|^2}\frac{h}{8e^2}\left(1+\exp\left(\frac{E_G}{k_BT} \right) \right)
\end{equation}

$\frac{h}{8e^2}$ denotes the quantum resistivity in graphene, $k_BT$ the thermal energy and $|t|^2$ the transmission probability of the tube. The ballistic transport regime only holds for ideal CNTs. The transport regime depends on the mean free path $\lambda$ of the electrons and the extension of the electronic wave function being infinite for ideal CNTs. Values for $\lambda$ of about several $\mathrm{\mu}$m are estimated theoretically in \cite{Sundqvist_2008} and therefore, ballistic transport should hold for the planned sensor. Experimental measurements, in contrast, show a significant increase of the resistance of CNTs at room temperature \cite{Purewal_2007} -- perhaps due to defects within them. Nevertheless, even if the results are not quantitatively correct, the change due to deformation is important and this can be estimated by this approach.

\begin{figure}[!t]
\centering
\includegraphics[width=\linewidth]{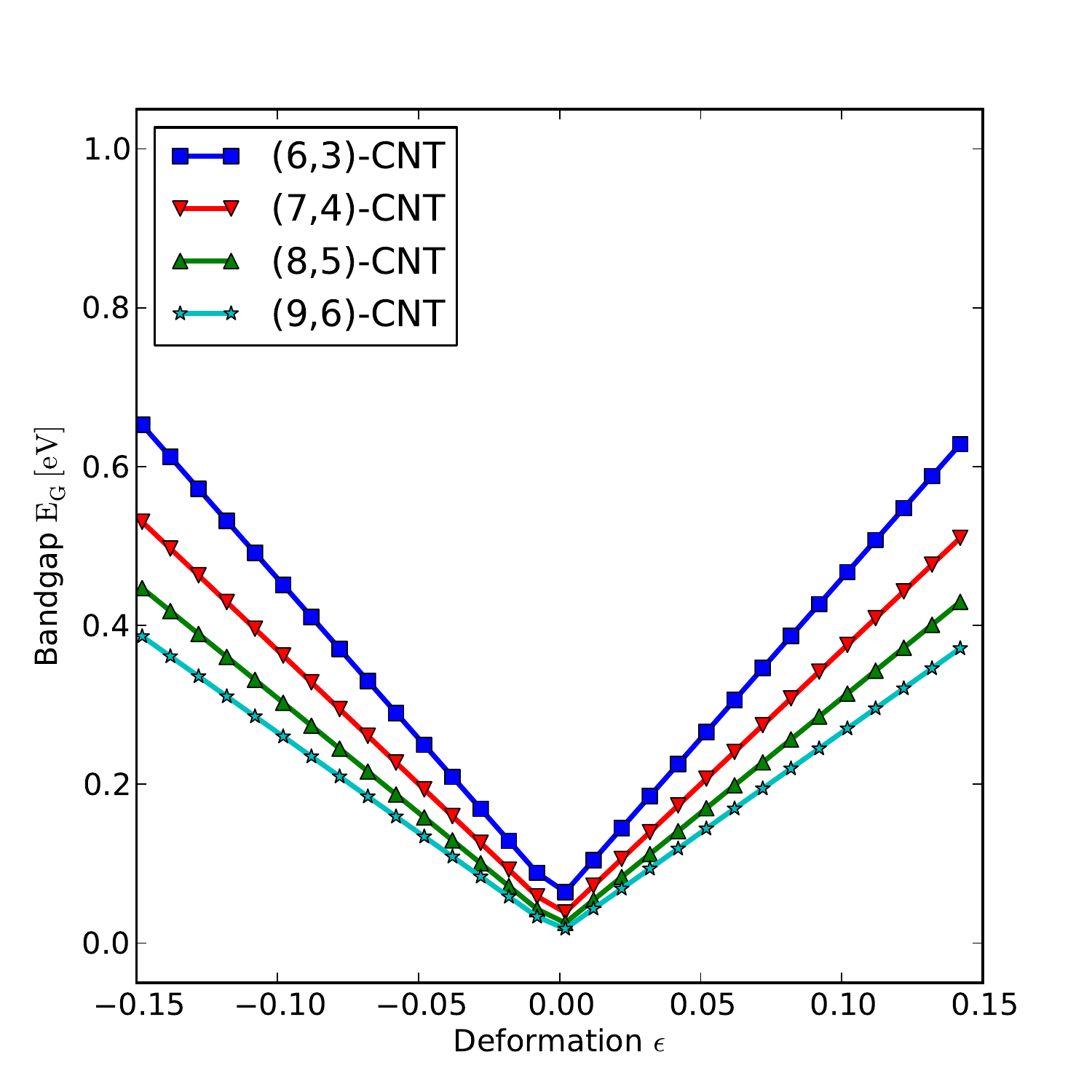}
\caption{The bandgap of CNTs over strain within the tight binding model is shown. The bandgap shows a linear dependence on strain in this model. The ascend of the bandgap is influenced by the chiral angle, monotonically increasing from (6,3) to (9,6) CNT.}
\label{fig_Bandgap}
\end{figure}

Thus, the information we need is the change of the bandgap of the system. The way to do this analytically is to linearly extrapolate the dispersion of the Dirac-(k-)-point in graphene. The CNT with its periodicity constraints of the unit cell introduces k-lines in the brillouin zone and forms the dispersion relation of a CNT. The deformation induced in the CNT is a coordinate transformation inducing relative shifting of the Dirac-point to the k-lines in the k-space. This is explicitely calculated in the publication of Yang and Han \cite{Yang_2000} and leads to linear changes of the bandgap over strain. Results of such calculations based on the deformation data shown before are presented in figure \ref{fig_Bandgap}, using a slightly different theory by Kleiner and Eggert adapted for semimetallic CNTs \cite{Kleiner_2001}. This theory shows a finite band gap at zero strain in contrast to the theory by Yang and Han.

So far, the tight binding model has been used for the prediction of the bandgap change. In  a future step, these results will be refined by DFT-calculations. Both theories predict a significant shift of the bandgap resulting in an exponential increase in the resistivity (see equation 2). 

\section{Conclusion}

DFT calculations of the mechanical properties of selected CNTs are reported which are consistent with the current literature. The force-distance curves deviates slightly from a linear elastic behavior for strains higher than 5\%. The total energy of the system can be nicely fit by a third order polynomial over a wide range of deformation so that the data can easily be condensed in a compact model for the mechanical behavior, needed for a sensor system device simulation. Values for the compact models for two relevant kind of CNTs are presented. 

The next step is the validation of  the tight binding model for application. If so, one could directly enter this model into the description of the sensor. First results indicate however that the bandgaps and the electronic behavior calculated by DFT show some significant deviations from the results predicted by tight binding. The preliminary data depends sensitively on the mechanical structure, whereas the total energy, determining the mechanical behavior, is robust. Thus, DFT using MD geometries cannot be used by DFT to calculate electronic structure, yet. This is a matter of ongoing investigations.

A further extension of the model to include biased transport simulations based on non-equilibrium Green's functions (NEGF) is underway and will be issue of forthcoming publications. 

\section*{Acknowledgement}

This work has been done within the VW foundation and the Research Unit 1713 which is funded by the German Research Association (DFG). We gratefully acknowledge the ongoing support by the group of Michael Schreiber (TU Chemnitz).

\IEEEtriggeratref{10}

\end{document}